\documentclass[%
 aip,
 jmp,%
 amsmath,amssymb,
preprint,%
]{revtex4-1}

\usepackage{graphicx}
\usepackage{dcolumn}
\usepackage{bm}
\usepackage{epstopdf}
\usepackage{amssymb}
\usepackage{color}
\usepackage[colorlinks=true, letterpaper=true, pdfstartview=FitV, linkcolor=blue, citecolor=blue, urlcolor=blue]{hyperref}
\usepackage{lineno}

\begin{document}

\title{Prediction of interesting ferromagnetism in Janus semiconducting Cr$_2$AsP monolayer}
\author{Qiuyue Ma}
\author{Yingmei Li}
\author{Guochun Yang}
\author{Yong Liu}\email{yongliu@ysu.edu.cn}
\affiliation{State Key Laboratory of Metastable Materials Science and Technology \& Key Laboratory for Microstructural Material Physics of Hebei Province, School of Science, Yanshan University, Qinhuangdao 066004, China }

\begin{abstract}
Two-dimensional (2D) half-metallic materials that have sparked intense interest in advanced spintronic applications are essential to the developing next-generation nanospintronic devices. Here we have adopted a first-principles calculation method to predict the magnetic properties of intrinsic, Se-doped, and biaxial strain tuning Cr$_2$AsP monolayer. The Janus Cr$_2$AsP monolayer is proven to be an intrinsic ferromagnetic (FM) semiconductor with an exchange splitting bandgap of 0.15 eV at the PBE+U level. Concentration-dependent Se doping such as Cr$_2$As$_{1-x}$Se$_x$P (x = 0.25, 0.50, 0.75) can regulate Cr$_2$AsP from FM semiconductor to FM half-metallicity. Specifically, the spin-up channel crosses the Fermi level, while the spin-down channel has a bandgap. More interestingly, the wide half-metallic bandgaps and spin bandgaps make them have important implications for the preparation of spintronic devices. At last, we also explore the effect of biaxial strain from -14\% to 10\% on the magnetism of the Cr$_2$AsP monolayer. There appears a transition from FM to antiferromagnetic (AFM) at a compressive strain of -10.7\%, originating from the competition between the indirect FM superexchange interaction and the direct AFM interaction between the nearest-neighbor Cr atoms. Additionally, when the compressive strain to -2\% or the tensile strain to 6\%, the semiconducting Cr$_2$AsP becomes a half-metallic  material. These charming properties render the Janus Cr$_2$AsP monolayer with great potential for applications in spintronic devices.

\end{abstract}

\maketitle


\maketitle

Spintronic device using the spin degree freedom of electrons has attracted tremendous interest over the past decades owing to its lower power consumption,  greater data processing speed, and higher integration densities~\cite{S. A.-Science-2001}. Two-dimensional (2D) magnetic materials provide new opportunities for spintronics and nanoscale magnetic memory devices. Spintronics magnetic materials need to have a high spin polarization rate~\cite{1S. A.-Science-2001}. In 1983, through the study of alloys such as Heusler alloys NiMnSb and PtMnSb, Groot et al. first discovered a material with a unique energy band structure, named a half-metallic ferromagnet~\cite{R. A-PRL-1963}, exhibiting a metallic property in one spin channel and an energy gap similar to a semiconductor in the other spin channel. Therefore, they possess 100\% spin polarization at the Fermi level, and are good source of spin-flow injection and can meet the demands of high-performance spintronic devices~\cite{S. M-PRB-2000,W.-H.-J-2004,B. Doumi-E-2015}. Thus far, many FM half-metals have been predicted by numerous studies, such as MnX (X= P, As)~\cite{27B. Wang-Nanoscale-2019}, FeX$_2$ (X = Cl, Br, I)~\cite{R. K-Phys. Rev. B-2021}, NbF$_3$~\cite{B. Yang-Phys-2018}, Cr$_2$NX$_2$ (X = O, F, OH)~\cite{Q. Sun-Appl-2021}, and FeXY (X, Y = Cl, Br, and I, X $\neq$ Y)~\cite{28R. Li-a-2021}.

Compared with traditional semiconductor devices, spintronic devices have superior performance. One of the practical routes to obtain the compatibility of electronic materials is the introduction of highly concentrated magnetic ions to make non-magnetic semiconductors magnetic, or even ferromagnetic transition. With the development of spintronics materials, new magnetic materials with both magnetic and semiconducting properties have been realized by injecting transition metal ions into a binary semiconductor, contributing to the progress of spintronics. In recent years, Fe-doped semiconductors have received much attention as ferromagnetic semiconductors because of their high Curie temperatures and low power consumption, showing the potential use in high-speed spin devices. High Curie temperature ferromagnetism was also observed in Fe-doped InAs, from which n-type and p-type ferromagnetic semiconductors can be prepared~\cite{H. Shinya-J-2018}. Furthermore, it has been found that the doping of small amounts of magnetic elements such as Group II-VI~\cite{H. Liu-p-2017,H.-F-s-2017}, IV, and III-V into semiconductors~\cite{J. Coey-S-2005,K. Yang-p-2010,H. Katayama-p-2003}. Specifically, the doped magnetic atoms replace cations or anions in the semiconductor unit cell, or the formation of defects in the studied system by defect techniques, which has led to the discovery of many new spintronics materials~\cite{N. T-p-2016,L. D-A-2015}. Recently, we have identified several half-metallic materials by using transition metal elements to dope group III-V binary semiconductors~\cite{21Y. Liu-J-2007,22N. Noor-J-2011,23G. Rahman-P-2010,24M. Shirai-P-2001,K.-W.-C-2016}. Se\~{z}a et al~\cite{N. Sea-j-2016}investigated Mn-doped GaSb using the density functional theory method and found that the doped material has ferromagnetic half-metallic properties. On the other hand, magnetic half-metallic materials made by doping have better compatibility compared with semiconductors and have shown high research and application value, so the research on this type of half-metallic materials have attracted increasing attention.

The modulation and control of spin ordering is a key issue for spintronic device applications. Mechanical strain is commonly considered to be an effective solution for regulating the electronic structure and magnetic properties of the 2D materials. The excellent mechanical flexibility of 2D magnets further demonstrates the feasibility of strain engineering~\cite{F. Li-n-2013,H. H-n-2014}. The application of tunable biaxial strain to 2D materials is of great significance for the preparation of spintronic devices. 2D materials are more flexible, and strains can be generated by external manipulations, such as bending and electric fields~\cite{J. Das-A-2009,Z. Liu-n-2014,S. Zhang-s-2017,R. Frisenda-n-2017,Y. Sun-j-2019}. For example, as the strain changes from 10\% to -15\%, the CrI$_3$ monolayer undergoes a transition from semiconductor to metal~\cite{Zewen-P-2019}. Particularly, an antiferromagnetic (AFM) to FM transition occurs under a biaxial tensile strain of approximately 13\% in MnPSe$_3$~\cite{Q. Pei-F-2018}. Experimentally, applying tunable biaxial strain to 2D materials has also made significant progress~\cite{F. Ding-N-2010}.

In this work, we investigated the electronic structures and the magnetic properties of the intrinsic, Se-doped and biaxial strain tuning Janus Cr$_2$AsP monolayer by the density functional theory calculations. The results indicated that the Janus Cr$_2$AsP monolayer is a FM semiconductor, which was consistent with previous study. After inducing substituted selenium (Se) dopants, Cr$_2$As$_{1-x}$Se$_x$P (x = 0.25, 0.50, 0.75) with wide bandgaps, show half-metallic ferromagnetism, indicating that Cr$_2$As$_{1-x}$Se$_x$P can be widely used in spintronic devices. Furthermore, we applied a biaxial strain range from -14\% to 10\%. When an approximately -10.7\% compressive strain was applied to the Cr$_2$AsP monolayer, it leads to a FM to AFM transition. Besides, the semiconductor of Cr$_2$AsP becomes half-matal within a certain tensile or compressive strain. These studies
first imply that the Janus Cr$_2$AsP monolayer is a potential spintronic material.

The present calculations were performed by adopting the Vienna ab initio simulation package (VASP) based on the density functional theory (DFT)~\cite{27P.-Phys. Rev. B-1994,28G. Kresse-American Physical Society-1996,29G. Kresse-Computational Materials Science-1996}. The generalized gradient approximation (GGA) functional of Perdew, Burke, and Ernzerhof (PBE) was used to investigate the exchange-correlation
function~\cite{30J. P. Perdew-Phys. Rev. Lett.-1996}. We used the spin-dependent GGA plus Hubbard U to deal with the strongly correlated interactions of the transition metal Cr element, the Hubbard U term of 3 eV for Cr was used~\cite{32A. I. Liechtenstein-American Physical Society-1995}. The plane-wave cutoff energy was chosen to be 500 eV. Monkhorst-Pack special k-point mesh of $9\times9\times1$ for the Brillouin zone integration~\cite{31H. J. Monkhorst-Phys. Rev. B-1976}. The convergence criteria for energy and force during the relaxation of the structures were set to 10$^{-6}$ eV and 0.01 eV/{\AA}. The vertical vacuum spacing of 20 {\AA} was used to eliminate interactions between images. Phonon dispersions of the studied materials were obtained using the phonopy code based on the density functional perturbation theory (DFPT)~\cite{33S. Baroni-Rev. Mod. Phys.-2001}.

\begin{figure}[t!hp]
\centerline{\includegraphics[width=0.9\textwidth]{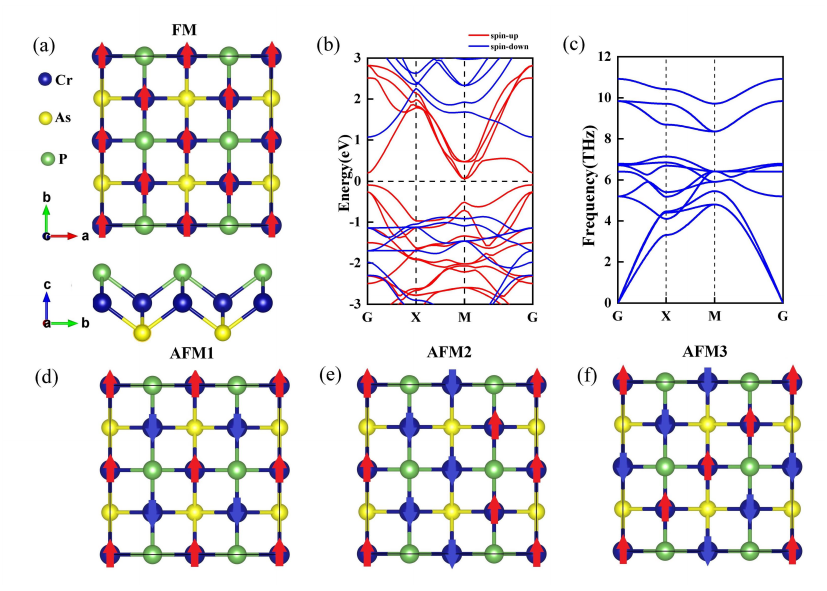}}
\caption{(a) Top and side views of Janus Cr$_2$AsP monolayer. (b)Band structure and (c) phonon spectra of Janus Cr$_2$AsP. Four magnetic configurations of Cr$_2$AsP: (a) FM, (d) AFM1, (e) AFM2, (f) AFM3.
\label{fig:1}}
\end{figure}


The top and side views of Janus Cr$_2$AsP monolayer are shown in Fig.~\ref{fig:1}(a), a sheet of Cr atoms is sandwiched between a sheet of As atoms and a sheet of P atoms. The optimized lattice constant is  a = b = 4.24 {\AA} with a Cr-Cr distance of 3.0 {\AA}. 2 $\times$ 1 $\times$ 1 supercell was used to calculate the magnetic ground state of Cr$_2$AsP monolayer, we selected one FM and three AFM configurations as shown in Fig.~\ref{fig:1}(a)(d)(e)(f). We found that the ground state of Janus Cr$_2$AsP monolayer was FM  by comparing the energy differences between the FM configuration and the AFM configurations. Furthermore, the magnetic moment of each primitive cell was about 6 $\mu_B$, and the local magnetic moment of a Cr atom was nearly 3.4 $\mu_B$. The band structure of Janus Cr$_2$AsP monolayer calculated by the PBE+U functional is shown in Fig.~\ref{fig:1}(b). The valence band maximum (VBM) is located at G points and the conduction band minimun (CBM) is located at M points, demonstrating that the Janus Cr$_2$AsP monolayer is an indirect band gap semiconductor, and the exchange splitting bandgap is 0.15 eV, a more accurate value is calculated as 0.70 eV by using  Heyd-Scuseria-Ernzerhof (HSE06) hybrid functional. Besides, we calculated the elastic constants, which clearly satisfy the Born criterion of stability~\cite{34Z.-Phys. Rev. B-2007}. And as shown in Fig.~\ref{fig:1}(c), we also calculated the phonon dispersion relation, and the imaginary frequency was found to be absent. These results indicate that the Janus Cr$_2$AsP monolayer is dynamically stable.

\begin{figure}[tbp!]
\centerline{\includegraphics[width=0.95\textwidth]{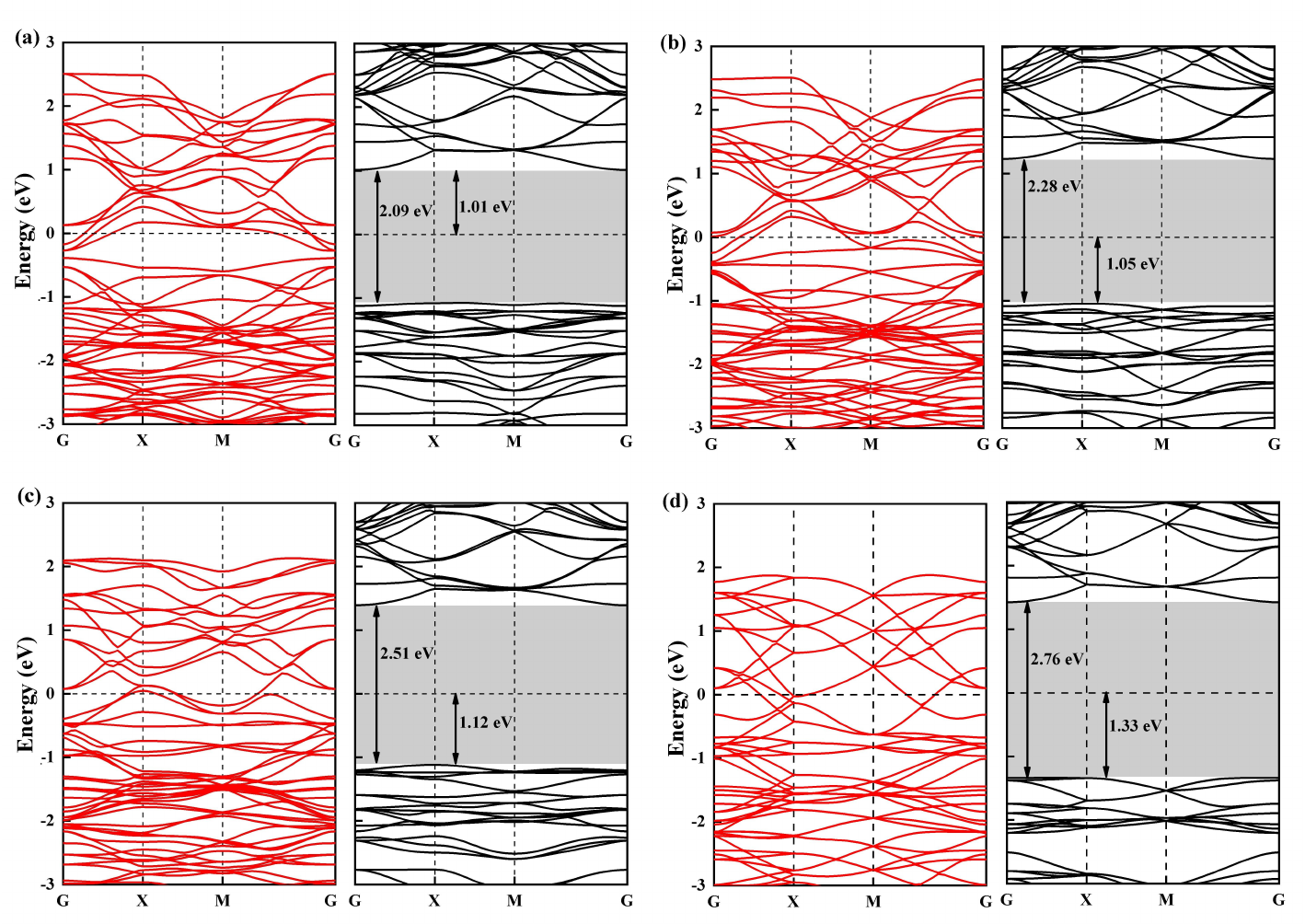}}
\caption{Band structures of (a) Cr$_2$As$_{0.25}$Se$_{0.75}$P; (b) Cr$_2$As$_{0.5}$Se$_{0.5}$P; (c) Cr$_2$As$_{0.75}$Se$_{0.25}$P; (d) Cr$_2$SeP monolayers.
\label{fig:2}}
\end{figure}

In the study of spintronic devices, it is difficult to avoid dopants or defects, but the purposeful induction of dopants or defects can regulate the properties of 2D materials, which plays an important role in facilitating the preparation of spintronic devices. In this article, we investigated the Se-doped Cr$_2$AsP monolayer. We used 2 $\times$ 1 $\times$ 1 supercell to consider the replacement of As atoms in Cr$_2$AsP monolayer by one, two, and three Se atoms, and the ion-injected Se ratios are 25\%, 50\%, and 75\%. We defined the x as the ion-injected Se ratios.

Table~\ref{table:elastic} shows the crystal parameters and properties of Cr$_2$As$_{1-x}$Se$_x$P (x = 0.25, 0.50, 0.75, 1.00), we can see that the equilibrium lattics constant for Cr$_2$As$_{0.25}$Se$_{0.75}$P, Cr$_2$As$_{0.5}$Se$_{0.5}$P, Cr$_2$As$_{0.75}$Se$_{0.25}$P, and Cr$_2$SeP monolayers are 8.48 {\AA}, 8.56 {\AA}, 8.71 {\AA}, and 8.85 {\AA}, respectively. Then, we calculated the ground state magnetic order of Cr$_2$As$_{1-x}$Se$_x$P (x = 0.25, 0.50, 0.75, 1.00) by comparing the energy difference between FM and AFM configurations, and found that the FM energy of the material after doping is lower, so all of Cr$_2$As$_{1-x}$Se$_x$P exhibit FM state. The electronic structures of Cr$_2$As$_{1-x}$Se$_x$P (x = 0.25, 0.50, 0.75, 1.00) were calculated in the FM ground state. As shown in Fig.~\ref{fig:2}, Cr$_2$As$_{1-x}$Se$_x$P (x = 0.25, 0.50, 0.75, 1.00) have similar energy band structures, the spin-up channel acrosses the Fermi level exhibiting metallicity, the spin-down channel has band gap at the Fermi level showing semiconductivity, which indicates that they are fully spin-polarized half-metals. After doping, all of Cr$_2$As$_{1-x}$Se$_x$P (x = 0.25, 0.5, 0.75, 1.00) exhibit large half-metallic gaps and spin gaps, wide half-metallic gaps and wide spin gaps are the key half-metal parameters in spintronic applications. Herein, the half-metallic gaps of the Cr$_2$As$_{0.25}$Se$_{0.75}$P, Cr$_2$As$_{0.5}$Se$_{0.5}$P, Cr$_2$As$_{0.75}$Se$_{0.25}$P, and Cr$_2$SeP monolayers are 1.01 eV, 1.05 eV, 1.12 eV, and 1.33 eV, respectively, which are larger than the reported in previous research on Mn$_2$PAs (0.68 eV)\cite{40H. Zeng-J. Mater. Sci.-2021}; the spin gaps are 2.09 eV, 2.28 eV, 2.51 eV, and 2.76 eV, respectively. It can be noticed that the half-metallic bandgaps and spin bandgaps increase with the increase of the proportion of injected Se atom.

To further investigate the rationale for the half-metallic nature of this series of half-metallic ferromagnets, the total density of states of Cr$_2$As$_{0.25}$Se$_{0.75}$P is selected because the electronic densities of the Cr$_2$As$_{1-x}$Se$_x$P (x = 0.25, 0.5, 0.75) are similar. Since the electronic structure and magnetic properties of most materials are derived from intermetallic orbital p-d, d-d, and s-p-d orbital hybridization, and the orbitals are also related to the electron configuration of each atom and the distance between the atoms. The valence electron configuration of Cr is 3$d^5$4$s^1$, the Se and As valence electron configurations are 4$s^2$4$p^4$ and 4$s^2$4$p^3$, the outermost 4p orbit has four electrons and three electrons in Se and As. And the valence electron configuration of P is 4$s^2$3$p^3$, the outermost 3p orbit has three electrons. By comparing the bond lengths and density of states diagrams among the atoms in the crystal, we believe that strong hybridization occurs between the electronic orbitals near the Fermi level and that the electronic hybridization energy occurs mainly between the transition metal Cr and As, Se , and P atoms. As shown in Fig.S1, supplement meterials, from the density of each atomic fractional wave state it can be seen that the total density of states for Cr$_2$As$_{0.25}$Se$_{0.75}$P is mainly contributed by Cr-d orbital electrons, while other electronic orbitals such as As-p, Se-p and P-p contribute less. We find that the spin-down states of Cr-d, As-p, and P-p have overlapping densities of states around the bandgaps, so that there is p-d orbital hybridization in the corresponding energy region, and a small amount of orbital hybridization between Cr-d and Se-p after doping with Se. The p-d hybridization from the Cr-3d state with the Se-4p state may lead to a shift in Fermi level, with the spin-down energy band of Cr$_2$As$_{0.25}$Se$_{0.75}$P forming a bandgap at the Fermi level. From the band structures that the spin gap values increase with doping concentration, and from this change, it is clear that the p-d hybridization between Cr-3d and Se-4p intensifies with increasing doping concentration, which is the main reason for the FM half-metallic properties of Cr$_2$As$_{1-x}$Se$_x$P. The Fermi level of Cr$_2$As$_{0.25}$Se$_{0.75}$P, Cr$_2$As$_{0.5}$Se$_{0.5}$P, and Cr$_2$As$_{0.75}$Se$_{0.25}$P is calculated to have shifted upwards by 0.25 eV, 0.22 eV, and 0.20 eV respectively when compared to the density of states of the Cr$_2$AsP monolayer. The doped Cr$_2$AsP is thus transformed from a semiconductor to a half-metallic material. Finally, the calculated phonon dispersions in Fig.S2 support their dynamical stability, supplement meterials.

\begin{table}[h]
\begin{ruledtabular}
\caption{Lattice constants a$_0$ ({\AA}), ground state magnetic properties (GP), material properties (MP) (the HMF indicates a half-metallic ferromagnet), half-metallic gap (HM) , semiconductor gap (SM) for Cr$_2$As$_{1-x}$Se$_x$P (x = 0.25, 0.50, 0.75, 1.00) monolayers.}\label{table2}
\begin{tabular}{lccccc}
   System & a$_0$ ({\AA}) &GP & MP & HM (eV)& SM (eV)  \\
   \hline

Cr$_2$As$_{0.25}$Se$_{0.75}$P  & 8.48 & FM & HMF & 1.01 &2.09\\

Cr$_2$As$_{0.5}$Se$_{0.5}$P &8.56 &FM & HMF & 1.05 & 2.28\\

Cr$_2$As$_{0.75}$Se$_{0.25}$P &8.71 &FM & HMF &1.12 &2.51\\

Cr$_2$SeP  &8.85 &FM & HMF &1.33  & 2.76
\label{table:elastic}
\end{tabular}
\end{ruledtabular}
\end{table}

\begin{figure}[tbp!]
\centerline{\includegraphics[width=0.95\textwidth]{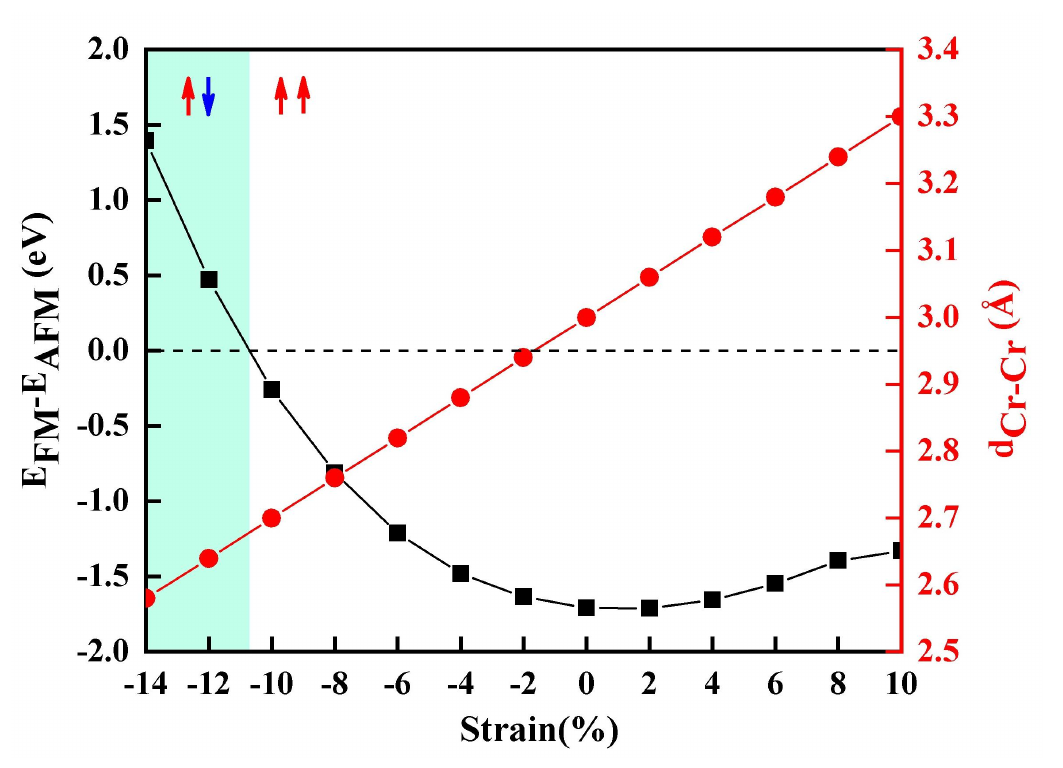}}
\caption{Energy difference between the FM and AFM phases (black line) and the nearest Cr-Cr distance (red line) under the biaxial strain for monolayer Cr$_2$AsP.
\label{fig:3}}
\end{figure}

\begin{figure*}[htp]
\centerline{\includegraphics[width=0.9\textwidth]{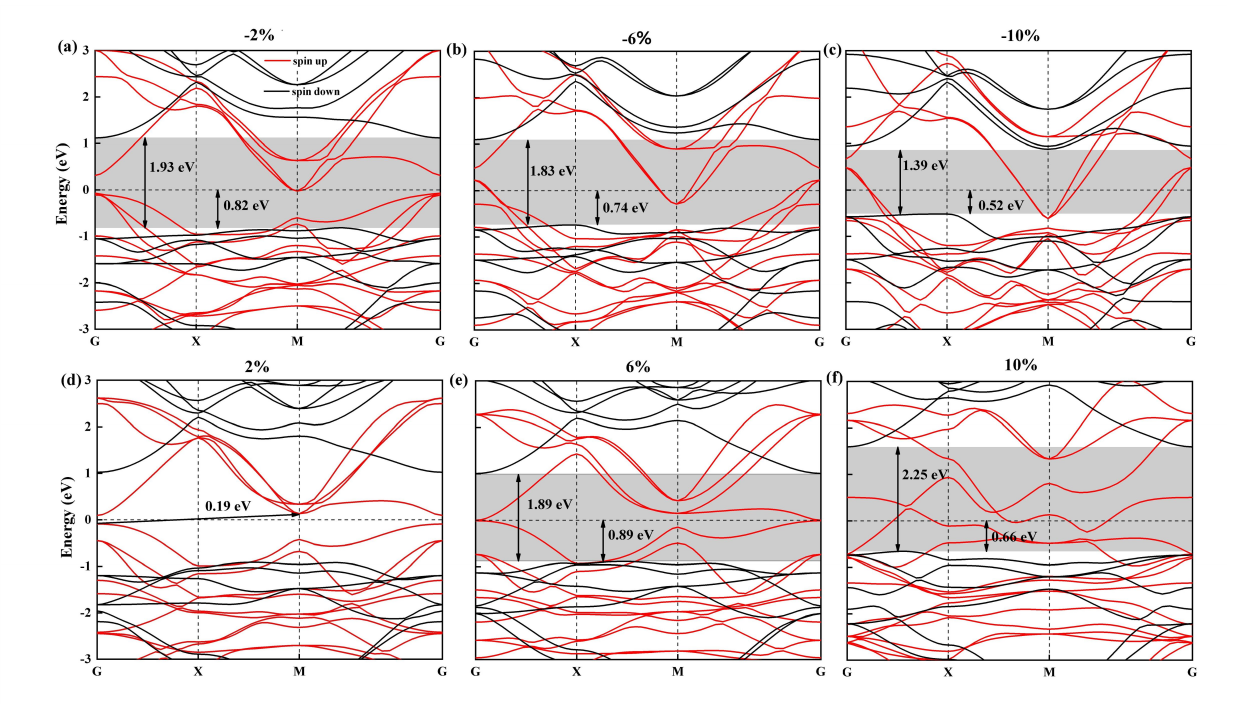}}
\caption{Band structures for Janus Cr$_2$AsP monolayer under the biaxial strain.(a) -2\%; (b) -6\%; (c) -10\%; (d)2\%; (e)6\%; (f)10\%.}
\label{fig:4}
\end{figure*}

Next, we investigated the electronic and magnetic properties of Janus Cr$_2$AsP monolayer under the biaxial strain. The strain is defined as $\varepsilon  = \frac{{(a - {a_0})}}{{{a_0}}}$ , where a$_0$ and a are the lattice constant for unstrained and strained system. The black line in Fig.~\ref{fig:3} shows the energy difference ($\Delta$E) between FM and AFM configurations. Within a certain range of tension and compression, the $\Delta$E becomes larger. A phase transition from FM to AFM occurs when a compressive strain achieves approximately -10.7\%. Large strain modulation has been observed in graphene~\cite{35T. Zhu-Progress in Materials Science-2010}. Therefore, we believe that the nearly -10.7\% compressive strain of the Cr$_2$AsP monolayer can be achieved. Moreover, the distance of the two nearest-neighbor Cr atom increases from 2.58 {\AA} to 3.30 {\AA} under the biaxial strain from -14\% to 10\%. The exchange interaction between atomic spins can be used to explain the FM to AFM transition of monolayer Cr$_2$AsP caused by elastic strain. We described the magnetic properties of Cr$_2$AsP by using the Heisenberg model Hamiltonian:

\begin{equation}
{H =  - \sum\limits_{ < ij > } {{J_{1}}{S_i}{S_j}}- \sum\limits_{ < ik > } {{J_{2}}{S_i}{S_k}}- A{{S_i^Z}{S_i^Z}}}
\end{equation}

J$_1$ and J$_2$ represent the exchange interactions between the nearest-neighbor and second nearest-neighbor spins. S$_i$ and S$_j$ represent the spin operators on site i and j, respectively; A is the parameter of magnetic anisotropy, and S$_Z$ is the spin along z direction. The energies of the different magnetic configurations are as follows:

\begin{equation}
{E(FM) = {E_0} - 2{J_1}{S^2}- 2{J_2}{S^2}- A{S^2}}\\
\end{equation}

\begin{equation}
{E(AFM1) = {E_0} + 2{J_1}{S^2}- 2{J_2}{S^2}- A{S^2}}\\
\end{equation}

\begin{equation}
{E(AFM3) = {E_0} + 2{J_2}{S^2}- A{S^2}}\\
\end{equation}

Here, S=3/2, the J$_1$ and J$_2$ are calculated as 36.07 and 25.80 meV per Cr atom, and the positive J value indicates the FM state. The J$_1$ mainly follows two kinds of exchange interactions. One is the direct exchange interactions among Cr-Cr, which is AFM for the high spin state of Cr$^{3+}$. The other is the superexchange interactions among Cr-As(P)-Cr. According to the Goodenough-Kanamori-Anderson (GKA) rules ~\cite{36J. B-Phys. Rev.-1955,37J. Kanamori-Journal of Applied Physics-1960}, the system prefers FM ordering when the cation-anion-cation bond angle is approximately 90$^\circ$ and AFM ordering when this angle is approximately 180$^\circ$. The bond angle of Cr-As-Cr (77.08$^\circ$) and Cr-P-Cr(72.63$^\circ$) in the Janus Cr$_2$AsP monolayer are both close to 90$^\circ$. Therefore, the Cr-As(P)-Cr angle in superexchange is FM. Due to the large distance between second nearest-neighbor Cr atoms, the superexchange interaction is dominant. A phase transition from FM to AFM occurs when a -10.7\% compressive strain is applied to the Cr$_2$AsP monolayer. Because the nearest neighbor Cr-Cr distance gets shorter, the direct exchange dominates over superexchange. When a tensile strain is applied to Cr$_2$AsP, the Cr-Cr distance will become larger, and the bond angle of Cr-As(P)-Cr will be more closer to 90$^\circ$, thus the system still exhibits FM state. As shown in Fig.~\ref{fig:4}, when compressive strain and tensile strain are added to -2\% and 6\%, it changes from semiconductor to half-metal. The half-metallic materials are conducting in one spin channel and insulating in another optional channel, which are 100\% spin polarized around the Fermi level~\cite{38R. A. de Groot-Phys. Rev. Lett.-1983,39A. Schindlmayr-New J. Phys.-2007}. The strained Cr$_2$AsP monolayer with half-metallicity has large spin bandgap which can effectively prevent spin leakage, becoming very important candidate materials in nanoscale spintronic devices~\cite{40C. Felser-Angewandte-2007}. Elastic strain engineering can find more materials with better properties to promote the application of spintronics.

In summary, we have investigated the electronic structures and the magnetic properties of intrinsic, Se-doped, and biaxial strain tuning in the Cr$_2$AsP monolayer by using the first-principles calculations. On the one hand, we found that the Cr$_2$AsP monolayer was FM semiconductor with an exchange splitting of 0.15 eV. When the Se doping ratio is 25\%, 50\%, and 75\%, the ground states of Cr$_2$As$_{1-x}$Se$_x$P (x = 0.25, 0.5, 0.75) were still FM. The Cr$_2$As$_{1-x}$Se$_x$P (x = 0.25, 0.5, 0.75) monolayers have the similar energy band structures, all exhibiting half-metallic properties. And Cr$_2$As$_{1-x}$Se$_x$P possess wide half-metallic bandgaps and spin bandgaps, which imply that they are more suitable for practical spintronic applications. On the other hand, a biaxial strain from -14\% to 10\% was applied to the Cr$_2$AsP monolayer. The phase transition from FM to AFM occurred at -10.7\% compressive strain for Cr$_2$AsP. When compressive strain was added to -2\% or a tensile strain was up to 6\%, the Janus Cr$_2$AsP monolayer undergoes a transiton from semiconductor to half-metallic material. These results will possess potential appliactions in spintronics and motivate further experimental studies.

Supplementary Material includes 2 figures.


This work was supported by the NSFC (No.21873017), the Innovation Capability Improvement Project of Hebei province (22567605H), the Natural Science Foundation of Hebei Province of China (No. B2021203030), the Science and Technology Project of Hebei Education Department(No. QN2023177). The numerical calculations in this paper have been done on the supercomputing system in the High Performance Computing Center of Yanshan University.

The data that support the findings of this study are available from the corresponding author upon reasonable request.


\bibliography{apssamp}

\end{document}